\documentclass[preprint,showpacs,preprintnumbers,amsmath,amssymb]{revtex4}

\usepackage{latexsym}
\usepackage{amstext}
\usepackage[normalem]{ulem}

\newcommand{\mb}[1]{ \mbox{\boldmath$#1$}}
\newcommand{\ds}{\displaystyle}
\newcommand{\beq}{\begin{eqnarray}}
\newcommand{\eeq}{\end{eqnarray}}
\newcommand{\beqq}{\begin{eqnarray*}}
\newcommand{\eeqq}{\end{eqnarray*}}
\newcommand{\p}{\partial}

\newcommand{\eps}{\varepsilon}
\newcommand{\x}{\mbox{\boldmath$x$}}

\newcommand{\y}{\mbox{\boldmath$y$}}
\newcommand{\z}{\mbox{\boldmath$z$}}
\font\bb=msbm10 at 12pt
\def\rR{\hbox{\bb R}}

\begin{document}
\preprint{APS/123-PRE}
\title{Narrow escape and leakage of Brownian particles\\}
 \author{A. Singer}
 \email{amits@math.princeton.edu}
 \affiliation{Department of Mathematics and PACM, Princeton University, Fine Hall, Washington Road, Princeton NJ 08544-1000
USA}
 \author{Z. Schuss}
 \email{schuss@post.tau.ac.il}
 \affiliation{Department of Mathematics, Tel-Aviv University, Tel-Aviv 69978, Israel}
 \author{D. Holcman}
 \email{david.holcman@weizmann.ac.il}
 \affiliation{Department of Mathematics, Weizmann Institute of Science, Rehovot 76100, Israel}
 \affiliation{Visiting address: D\'epartement de Math\'ematiques et de Biologie, Ecole Normale Sup\'erieure, 46 rue d'Ulm 75005 Paris, France.}

\date{\today}

\begin{abstract}
Questions of flux regulation in biological cells raise renewed
interest in the narrow escape problem. The often inadequate
expansions of the narrow escape time are due to a not so well known
fact that the boundary singularity of Green's function for Poisson's
equation with Neumann and mixed Dirichlet-Neumann boundary
conditions in three-dimensions contains a logarithmic singularity.
Using this fact, we find the second term in the expansion of the
narrow escape time and in the expansion of the principal eigenvalue
of the Laplace equation with mixed Dirichlet-Neumann boundary
conditions, with small Dirichlet and large Neumann parts. We also
find the leakage flux of Brownian particles that diffuse from a
source to an absorbing target on a reflecting boundary of a domain,
if a small perforation is made in the reflecting boundary.
\end{abstract}


\maketitle

\section{Introduction}
The narrow escape problem in diffusion theory, which goes back to
Lord Rayleigh \cite{Rayleigh}, is to calculate the mean first
passage time, also called the narrow escape time (NET), of a
Brownian particle to a small absorbing window on the otherwise
reflecting boundary of a bounded domain. The renewed interest in the
small hole problem is due to its relevance in molecular biology and
biophysics. The small hole often represents a small target on a
cellular membrane, such as a protein channel, which is a target for
ions \cite{Hille}, a receptor for neurotransmitter molecules in a
neuronal synapse \cite{Nicoll}, a narrow neck in the neuronal spine,
which is a target for calcium ions \cite{EJN}, and so on. The
physiological role of the small hole is often to regulate flux,
which carries a physiological signal. For example, the NMDA channels
in the post synaptic membrane in the neuronal cleft are small
targets for diffusing glutamate molecules released from a vesicle at
the pre synaptic membrane. The leakage problem here is to find the
fraction of the released molecules that reach the channels before
being irreversibly absorbed by the surrounding medium (e.g., glia
transporters) \cite{Patton}, \cite{Gallo} (see also
http://en.wikipedia.org/wiki/Chemical\_synapse). The position and
the number of the NMDA and  AMPA receptors regulate synaptic
transmission and is believed to be a part of coding memory
\cite{Nicoll}, \cite{Malinow}.

The narrow escape problem is connected to that of calculating the
principal eigenvalue of the mixed Dirichlet-Neumann problem for the
Laplace equation in a bounded domain, whose Dirichlet boundary is
only a small patch on the otherwise Neumann boundary. Specifically,
the principal eigenvalue is asymptotically the reciprocal of the
narrow escape time in the limit of shrinking patch.

The recent history of the problem begins with the work of Ward,
Keller, Henshaw, Van De Velde, Kolokolnikov, and Titworth
\cite{Ward1,Ward2,Ward3,Ward4} on the principal eigenvalue and is
based on boundary layer theory and matched asymptotics, in which the
boundary layer equation is the classical electrified disk problem,
solved explicitly by Weber in 1873 \cite{Weber,Jackson}. The work of
Holcman, Singer, Schuss, and Eisenberg
\cite{HS,SSH1,SSH2,SSH3,PNAS,PLA-holes,JPA-holes,NarrowEscape4} on
the NET for diffusion with and without a force field and for several
small windows and its applications in biology, is based on the known
structure of the singularity of Neumann's function at the boundary
\cite{Jackson,CourantHilbert,Kellog,Garabedian} and on the Helmholtz
integral equation \cite{Helmholtz} (see \cite{Lurie}). The most
recent work of B\'enichou and Voituriez \cite{Benichou} on the NET
in diffusion and anomalous diffusion finds the dependence of the NET
on the initial point inside the boundary layer and finds the scaling
laws for sub-diffusions. In these papers the leading term in the
asymptotic expansion was calculated in the shrinking window limit.

Neither the second term, nor its order of magnitude were calculated
for the three dimensional problem, except in the case of a spherical
domain with a small circular absorbing window, where an explicit
solution was constructed by a generalization of Collins' method (an
error in the coefficient of the second term, given in \cite{SSH1},
is corrected here). The difficulty in finding, or even estimating,
the second term can be attributed to the practically unknown (to
mathematicians and physicists) structure of the singularity of
Neumann's function on the boundary. While classical texts in partial
differential equations and in classical mathematical physics
\cite{Jackson,CourantHilbert,Kellog,Garabedian} mention only the
leading order singularity of the Newtonian potential and a regular
correction, \cite{Kellog} shows (in an exercise) that Neumann's
function for a sphere has a logarithmic singularity at the boundary.
The logarithmic boundary singularity of Neumann's function for the
Laplace equation in a general regular domain seems to have been
discovered by Popov \cite{Popov} and elaborated by Silbergleit,
Mandel, and Nemenman \cite{Silbergleit} (which cites neither
\cite{Kellog} nor \cite{Popov}).

Another small window problem is that of a leaky conductor of
Brownian particles, which is a bounded domain with a source of
particles on the boundary or in the interior, and a (big) target,
which is an absorbing part of the boundary. The remaining boundary
is reflecting. If the boundary has a small absorbing patch (a hole),
some of the Brownian particles may leak out and never make it to the
big absorbing target. The calculation of the leakage flux is not the
same as that in the narrow escape problem, because the total flux on
the boundary remains bounded as the small hole shrinks. The
calculation of the leakage flux was attempted in \cite{Savtchenko}
for diffusion in a flat cylinder with a source at the reflecting top
and a small absorbing window at the reflecting bottom, and absorbing
lateral envelop. The three-dimensional diffusion in the cylinder was
assumed to be well approximated by radial diffusion in a circular
disk.

In this paper, we find the structure of the boundary singularity of
the Neumann function for the Poisson equation and of the
Green-Neumann function for the mixed problem (with Dirichlet and
Neumann boundary conditions) in a general bounded domain $\Omega$,
whose boundary $\p\Omega$ is sufficiently smooth. Our calculations
use the method of \cite{Popov,Silbergleit}. We find that for
$\z\in\p\Omega,\ \y\in\Omega\cup\p\Omega$, the structure of the
Neumann function (in dimensionless variables) is
 \beq
N(\y,\z)=\frac{1}{2\pi|\y-\z|}-\frac{1}{8\pi} \left[L(\z) +
N(\z)\right]\ln {|\y-\z|} + v_S(\y,\z),\label{Intro-v0z}
  \eeq
where $L(\z)$ and $N(\z)$ are the principal curvatures of $\p\Omega$
at $\z$ and $v_S(\y,\z)$ is a bounded function of $\x,\y$ in
$\Omega$. If $\Omega$ is a ball of radius $R$, the above mentioned
result of Kellog \cite{Kellog} is recovered, because
$L(\z)=N(\z)=\ds\frac1R$.

We find that the NET through a circular disk of (dimensionless)
radius $a$, centered at $\mb{0}$ on the boundary, is
 \beq
 E\tau=\frac{|\Omega|}{4aD\left[1+\ds\frac{L(\mb{0})+
N(\mb{0})}{2\pi}\,a\log a+o(a\log a)\right]},
 \eeq
where $D$ is the diffusion coefficient. If $\Omega$ is a ball of
radius $R$, then
 \beq
 E\tau=\ds{\frac{|\Omega|}{4aD}} \left[1+\frac{a}{\pi R}
\log \frac{R}{a} + o\left(\frac{a}{R}\log\frac{R}{a}\right)
\right].\label{Intro-Etau}
 \eeq
The result (\ref{Intro-Etau}) corrects that given in \cite{SSH1}.
The case of an elliptic window is handled in a straightforward
manner, as in \cite{SSH1}.

The principal eigenvalue $\lambda_1(a)\sim\ds\frac1{E\tau}$ of the
Laplace equation in $\Omega$ with Dirichlet conditions given on a
circular disk of dimensionless radius $a$ and Neumann boundary
conditions elsewhere has the asymptotic expansion
 \beq
 \lambda_1(a)=\frac{4aD}{|\Omega|}\left[1+\ds\frac{L(\mb{0})+
N(\mb{0})}{2\pi}\,a\log a+o(a\log a)\right]\quad\mbox{for}\quad
a\to0.\label{Intro-lambda1}
 \eeq
The result (\ref{Intro-lambda1}) provides the missing second term
and estimate of the remainder, which was not given in
\cite{Ward1,Ward2,Ward3,Ward4,SSH1}.

For a leaky conductor, we find that the leakage flux through a
circular hole of small (dimensionless) radius $\eps$, centered at
$\mb{0}$, is
 \beq
 J_{\eps}=4\eps
Du_0(\mb{0})\left(1+O(\eps\log\eps)\right),\label{Intro-Jeps4}
 \eeq
where $u_0(\mb{0})$ is the solution of the reduced problem (without
the leak) at the hole.

Equation (\ref{Intro-Jeps4}) can be viewed as a generalization of
(\ref{Intro-lambda1}) in the sense that the factor $|\Omega|^{-1}$
in (\ref{Intro-lambda1}) can be interpreted as the uniform
concentration of the Brownian particle in $\Omega$. The uniform
concentration is the solution of the stationary diffusion equation
problem with Neumann conditions on the entire boundary, which is the
reduced problem for narrow escape. Thus the concentration
$u_0(\mb{\x})$ is a generalization of the fixed concentration
$|\Omega|^{-1}$ in (\ref{Intro-lambda1}).
\section{The singularity of Neumann's function}

Consider a bounded domain $\Omega\subset\rR^3$, given by
$\Omega=\{\x\in\rR^3\,:\,F(\x)<0\}$, where $F\in C^3(\rR^3)$. Our
purpose is to determine the singularity of Green's function for the
Laplace equation in $\Omega$ with Neumann boundary conditions
(called Neumann's function) and of Green's function for the mixed
Dirichlet and Neumann boundary conditions.

The Neumann function $N(\x,\y)$ for this domain is the solution of
the boundary value problem
 \beq
\Delta_{\x} N(\x,\y) &=& -\delta(\x-\y) + \frac{1}{|\Omega|},
\quad\mbox{for}\quad \x,\y\in \Omega
\label{Poisson} \\
&&\nonumber\\
\frac{\p N(\x,\y)}{\p \nu_{\x}} &=& 0,\quad\mbox{for} \quad
\x\in\p\Omega,\ \y \in \Omega,\label{NeumannBC}
 \eeq
where $\mb{\nu}(\x)$ is the outer unit normal to the boundary
$\partial \Omega$. If $\x$ or $\y$ (or both) are in $\partial
\Omega$, then only a half of any sufficiently small ball about a
boundary point is contained in $\Omega$, which means that the
singularity of Neumann's function is $\ds\frac1{2\pi|\x-\y|}$.
Therefore Neumann's function for $\y\in\p\Omega$ is written as
\begin{equation}
N(\x,\y) = \frac{1}{2\pi |\x-\y|} + v(\x,\y),\label{Nv}
\end{equation}
where $v(\x,\y)$ satisfies
\begin{equation}
\Delta_{\x} v(\x,\y) =
\frac{1}{|\Omega|}\quad\mbox{for}\quad\x\in\Omega,\ \y\in\p\Omega
\end{equation}
and the boundary condition
\begin{equation}
\frac{\partial v(\x,\y)}{\partial \nu_{\x}} = \frac{1}{2\pi}
\frac{\mb{\nu}(\x)\cdot
(\x-\y)}{|\x-\y|^3}\quad\mbox{for}\quad\x,\y\in\p\Omega.
\end{equation}
Green's identity requires the evaluation of two integrals. The first
is the volume integral, which by (\ref{Poisson}) is
 \beqq
 \int_{\Omega} \left[N(\x,\y) \Delta_{\x} v(\x,\z) - v(\x,\z)
\Delta_{\x} N(\x,\y) \right]\,d\x &=& \int_{\Omega} N(\x,\y)
\frac{1}{|\Omega|}\,d\x + v(\y,\z) \nonumber\\
&&\nonumber\\
&&- \frac{1}{|\Omega|}\int_{\Omega} v(\x,\z)\,d\x,
 \eeqq
and the second is the surface integral, which by (\ref{NeumannBC})
is
 \beqq
&&\oint_{\partial \Omega}\left[N(\x,\y) \frac{\partial
v(\x,\z)}{\partial \nu_{\x}} - v(\x,\z) \frac{\partial
N(\x,\y)}{\partial \nu_{\x}}\right]\,dS_{\x} =\\
&&\\
&& \oint_{\partial \Omega} \left[\frac{1}{2\pi|\x-\y|} + v(\x,\y)
\right]\frac{\mb{\nu}(\x)\cdot(\x-\z)}{2\pi |\x-\z|^3}\,dS_{\x}.
 \eeqq
Thus, for $\z\in\p\Omega$ Green's identity gives
 \beq
v(\y,\z) = &-&\frac{1}{|\Omega|}
\int_{\Omega}\left[N(\x,\y)-v(\x,\z)\right]\,d\x\nonumber\\
&&\nonumber\\
& +& \oint_{\partial \Omega} \left[\frac{1}{2\pi|\x-\y|} + v(\x,\y)
\right] \frac{\mb{\nu}(\x)\cdot (\x-\z)}{2\pi
|\x-\z|^3}\,dS_{\x}.\label{Green'sIdent}
 \eeq

To determine the singularity of this integral when $\y$ approaches
$\z$, we use the method of successive approximations to expand
$v(\x,\y)$ as
\begin{equation}
v(\x,\y) \sim v_0(\x,\y) + v_1(\x,\y) + v_2(\x,\y) + \ldots,
\end{equation}
where $v_{i+1}(\x,\y)$ is more regular than $v_i(\x,\y)$ (see
\cite{Silbergleit}). For $\y$ or $\z$ (or both) in $\p\Omega$, the
first term is the most singular part
\begin{equation}
v_0(\y,\z) = \frac{1}{4\pi^2} \oint_{\partial \Omega}
\frac{\mb{\nu}(\x)\cdot (\x-\z)}{|\x-\y||\x-\z|^3}\,dS_{\x}.
\end{equation}
To extract its dominant part, we reproduce here, for completeness,
the analysis of \cite{Popov} with only minor modifications. We
consider $\z\in
\partial \Omega$ and assume that the boundary near $\z$ is
sufficiently smooth. Moving the origin to $\z$, we set $\z=\mb{0}$.
Taking a sufficiently small patch $\partial \Omega_{\z}$ about $\z$, we assume
that it can be projected orthogonally onto a circular disk $D_a$ of
radius $a$ in the tangent plane to $\p\Omega$ at $\z$. We can
assume, therefore, that $\p\Omega_{\z}$ can be represented as
 \beq
 x_3=f_{\z}(x_1,x_2)=\frac12L(\z)x_1^2+\frac12N(\z)x_2^2+o(x_1^2+x_2^2)\quad\mbox{for}\quad(x_1,x_2)\in
 D_a.\label{canonical}
 \eeq
If $a$ is sufficiently small, then $o(x_1^2+x_2^2)\ll
Lx_1^2+Nx_2^2$. This canonical representation (\ref{canonical})
assumes that $\p\Omega_{\z}$ has at least one non-zero curvature and
that the quadratic part in Taylor's expansion of $f(x_1,x_2)$ about
the origin is represented in principal axes.

The asymptotically dominant part as $\y\to\z$ is determined by the
integral over the patch $\p\Omega_{\z}$, which we write as
\begin{equation}
v_0(\y,\mb{0}) \sim \frac{1}{4\pi^2} \int_{\partial \Omega_{\z}}
\frac{\mb{\nu}(\x) \cdot
\x\,dS_{\x}}{\sqrt{(x_1-y_1)^2+(x_2-y_2)^2+(x_3-y_3)^2}\left[x_1^2 +
x_2^2 + x_3^2 \right]^{3/2}}
\end{equation}
In the representation (\ref{canonical})
 \beqq
\mb{\nu}(\x) &=& \frac{(L(\z)x_1, N(\z)x_2, -1) +
o(\sqrt{x_1^2+x_2^2})}{\sqrt{1+L^2(\z)x_1^2+N^2(\z)x_2^2}}\\
&&\\
\mb{\nu}(\x) \cdot \x &=& \frac{L(\z)x_1^2 + N(\z)x_2^2 -
x_3}{\sqrt{1+L^2(\z)x_1^2+N^2(\z)x_2^2}}\\
&&\\
dS_{\x} &=& \sqrt{1+|\nabla f_{\z}|^2}\,dx_1\, dx_2 \sim \sqrt{1+L^2(\z)
x_1^2 + N^2(\z) x_2^2}\,\,dx_1\,dx_2,
 \eeqq
so that
\begin{equation}
v_0(\y,\mb{0}) \sim \frac{1}{4\pi^2} \int_{D_a}
\frac{\left(L(\z)x_1^2 + N(\z)x_2^2 -
x_3\right)\,dx_1\,dx_2}{\sqrt{(x_1-y_1)^2+(x_2-y_2)^2+(x_3-y_3)^2}\left[x_1^2
+ x_2^2 + x_3^2 \right]^{3/2}}.\label{vpatch}
\end{equation}

The patch $\p\Omega_{\z}$ is represented in polar coordinates in
$D_a$ as
\begin{equation}
(x_1,x_2,x_3) = \left(r\cos\phi,r\sin\phi,
r^2\left(\frac{L(\z)}{2}\cos^2\phi + \frac{N(\z)}{2}\sin^2 \phi +
o(1)\right)\right),
\end{equation}
so transforming $\y$ into spherical coordinates
\begin{eqnarray*}
(y_1,y_2,y_3)=|\y|( \sin \theta \cos \phi_0,\,\sin \theta \sin
\phi_0,\, \cos \theta).
\end{eqnarray*}
we can write (\ref{vpatch}) as
 \beq
v_0(\y,\mb{0}) &\sim& \frac{1}{4\pi^2}
\int_0^{2\pi}I(|\y|,\phi,\theta)\,d\phi,\label{Iphi}
 \eeq
where
 \beq
&&I(|\y|,\phi,\theta)=\int_0^a \frac{\left[\frac{1}{2}Lr^2 \cos^2
\phi + \frac{1}{2}Nr^2 \sin^2 \phi + o(r^2)\right] r\,dr}{\left[r^2
+ |\y|^2 - 2r|\y|\sin\theta \cos(\phi-\phi_0) +
O(r^2|\y|+r^4)\right]^{1/2} \left[r^2 + O(r^4) \right]^{3/2}} \nonumber \\
&&\nonumber\\
&\sim&\frac{1}{2} \left[L \cos^2 \phi + N \sin^2 \phi
\right]\int_0^a \frac{dr}{\left[r^2 + |\y|^2 - 2r|\y|\sin\theta
\cos(\phi-\phi_0) \right]^{1/2}}.\label{Popov}
 \eeq
Integration with respect to $r$ gives
 \beqq
&&\int_0^a \frac{dr}{\left[r^2 + |\y|^2 - 2r|\y|\sin\theta
\cos(\phi-\phi_0) \right]^{1/2}}=\\
&&\\
&&  \ln \frac{a - |\y|\sin\theta \cos(\phi-\phi_0) +
\sqrt{a^2+|\y|^2 - 2a|\y|\sin\theta \cos(\phi-\phi_0)}}{|\y|
\left(1-\sin\theta\cos (\phi-\phi_0) \right)}=\\
&&\\\
&&\ln \frac1{|\y|}+O(1),
 \eeqq
for $\y\neq\mb{0}$. It follows from (\ref{Iphi})  that for
$\y\neq\z$ the leading order singularity is
\begin{equation}
v_0(\y,\z) \sim \frac{1}{8\pi}  (L(\z) + N(\z))\ln \frac{1}{|\y-\z|}
+ O(1).\label{v0z}
\end{equation}
For further analysis of the $O(1)$ term, see \cite{Silbergleit}.

The canonical representation (\ref{canonical}) of a hemisphere of
(dimensionless) radius $R$ at the south pole is
$x_3=R-\sqrt{R^2-(x_1^2+x_2^2)}$, so $L(\z)=N(\z)=\ds\frac{1}{R}$.
Therefore, for $|\z|=R$,
\begin{equation}
N(\y,\z)=\frac{1}{2\pi|\y-\z|}+\frac{1}{4\pi R}\ln\frac1{ |\y-\z|} +
O(1),
\end{equation}
in agreement with \cite[p.247, Exercise 4]{Kellog}.

\section{Application to the narrow escape problem}

\subsection{Escape through a small circular
hole}\label{ss:small-circular}

As mentioned in the Introduction, the narrow escape problem
\cite{Ward1,Ward2,Ward3,Ward4,HS,SSH1,SSH2,SSH3,PNAS} is to
calculate the mean escape time of a Brownian particle from a bounded
domain $\Omega$, whose boundary is reflecting, except for a small
absorbing patch (or patches \cite{PLA-holes,JPA-holes})
$\p\Omega_a$. We assume here that $\p\Omega_a$ is a circular disk of
radius $a\ll|\Omega|^{1/3}$ and that a ball of radius $R\gg a$ can
be rolled on $\p\Omega$ inside $\Omega$. This means that there are
no narrow passages in $\Omega$. We denote
$\p\Omega_r=\p\Omega-\p\Omega_a$ and $\eps=a/|\Omega|^{1/3}$ and
investigate the limit $\eps\to0$. We assume that all coordinates
have been scaled with $|\Omega|^{1/3}$, so that all variables and
parameters are dimensionless.

The MFPT $u(\x)$ from a point $\x\in\Omega$ to $\p\Omega_a$ is the
solution of the mixed boundary value problem
\begin{eqnarray}
\Delta u(\mb{x}) &=& -\frac{1}{D}, \quad\mbox{for}\quad\x\in \Omega
\label{eq:v-general}\\
&&\nonumber\\
 u(\x) &=& 0 \quad\mbox{for}\quad\x\in
\partial\Omega_a\label{eq:boundary-condition-v} \\
&&\nonumber\\
 \frac{\partial u(\x)}{\p\nu_{\x}} &=& 0
\quad\mbox{for}\quad\x\in
\partial \Omega_r, \label{dvdn}
\end{eqnarray}
where $D$ is the diffusion coefficient. The compatibility condition,
\begin{equation}
\label{eq:comp} \int_{\partial \Omega_a} \frac{\partial
u(\x)}{\p\nu_{\x}}\,dS_{\x} =- \frac{|\Omega|}{D},
\end{equation}
is obtained by integrating (\ref{eq:v-general}) over $\Omega$ and
using  (\ref{eq:boundary-condition-v}) and (\ref{dvdn}).

Green's identity and the boundary conditions (\ref{NeumannBC}),
(\ref{eq:boundary-condition-v}), and (\ref{dvdn}) give
 \beq
u(\y) - \frac{1}{D} \int_{\Omega} N(\x ,\y) \,d\x  = \int_{\partial
\Omega} N(\x,\y) \frac{\partial u(\x )}{\p\nu}\,dS_{\x}
+C,\label{vx}
 \eeq
where
 \beq
 C=\frac{1}{|\Omega|}
\int_{
\Omega} u(\x) \,d{\x}.\label{C}
 \eeq
Following the argument in \cite{SSH1}, we note that $N(\x,\y)$ is an
integrable function independent of $\p\Omega_a$, whose integral is
uniformly bounded, whereas $C\to\infty$ as  $\eps\to0$. Setting
$g(\x)=\ds\frac{\partial u(\x)}{\p\nu_{\x}}$ for $\x\in\p\Omega_a$
and using the boundary condition (\ref{eq:boundary-condition-v}), we
obtain from (\ref{vx}) the integral equation for the flux density
$g(\x)$ in $\p\Omega_a$,
 \beq
 \int_{\p\Omega_a}N(\x,\y)g(\x)\,dS_{\x}=-C\quad\mbox{for}\quad\y\in
 \p\Omega_a,
  \eeq
which, in view of (\ref{Nv}), (\ref{v0z})  now becomes the
generalized Helmholtz equation \cite{Helmholtz}, \cite{SSH1}
 \beq
&&
\int_{\p\Omega_a}g(\x)\left[\frac{1}{2\pi|\x-\y|}+H(\x,\y)\log|\x-\y|+O(1)\right]\,dS_{\x}=-C\quad\mbox{for}\quad\y\in
 \p\Omega_a,\nonumber\\
 &&\label{Hxy}\\
&& H(\x,\y)= - \frac{1}{8\pi}  [L(\y) + N(\y)]\sim - \frac{1}{8\pi}
[L(\mb{0}) + N(\mb{0})],\quad\mbox{for}\quad\x,\y\in\p\Omega_a
\quad\mbox{for}\quad \eps\to0,\nonumber
 \eeq
where $L(\mb{0}), N(\mb{0})$ are the principal curvatures at the
center $\mb{0}$ of $\p\Omega_a$. To solve (\ref{Hxy}), we expand
$g(\x)=g_0(\x)+g_1(\x)+g_2(\x)+\cdots$, where $g_{i+1}(\x)\ll
g_i(\x)$ for $\eps\to0$ and choose
 \beq
 g_0(\x)=\ds\frac{-2C}{a\pi\sqrt{1-\ds\frac{|\x|^2}{a^2}}}.\label{g0}
  \eeq
It was shown in \cite{Rayleigh}, \cite{Lurie}, \cite{SSH1} that if
$\p\Omega_a$ is a circular disk of radius $a$, then
\begin{equation}
\frac{1}{2\pi}\int_{\p\Omega_a} \frac{g_0(\x )}{|\x -\y |}\,dS_{\x}
= C\quad\mbox{for all}\quad\y\in\p\Omega_a.\label{intg0}
\end{equation}
It follows that $g_1(\x)$ satisfies the integral equation
\begin{equation}
\frac{1}{2\pi}\int_{\p\Omega_a} \frac{g_1(\x )}{|\x -\y |}\,dS_{\x}
=\frac{2C}{a\pi}\int_{\p\Omega_a}\frac{H(\x,\y)\log|\x-\y|}{\sqrt{1-\ds\frac{|\x|^2}{a^2}}}\,dS_{\x}.\label{g1}
\end{equation}
Setting $y=a\mb{\eta},\,\x=a\mb{\xi}$, and changing to polar
coordinates in the integral on the right hand side of (\ref{g1}), we
obtain
\begin{equation}
\frac{1}{2\pi}\int_{\p\Omega_a} \frac{g_1(\x )}{|\x -\y |}\,dS_{\x}
=\frac{2Ca^2}{a\pi}\int_0^{2\pi}d\theta\int_0^1\frac{H(a\mb{\xi},a\mb{\eta})\left[\log
a+\log|\mb{\xi}-\mb{\eta}|\right]}{\sqrt{1-r^2}}\,r\,dr,
\end{equation}
which gives in the limit $\eps\to0$ (e.g., keeping $|\Omega|$ fixed
and $a\to0$) that
\begin{equation}
\frac{1}{2\pi}\int_{\p\Omega_a} \frac{g_1(\x )}{|\x -\y |}\,dS_{\x}
=-\frac{ C[L(\mb{0})+ N(\mb{0})]}{2\pi}\,a\log a+o(a\log a).
\end{equation}
As in the pair (\ref{g0}), (\ref{intg0}), we obtain that
 \beq
 g_1(\x)=\frac{-C[L(\mb{0})+
 N(\mb{0})]}{\pi^2\sqrt{1-\ds\frac{|\x|^2}{a^2}}}\,\log a+o(\log a).
 \eeq
Finally, to determine the asymptotic value of the constant $C$, we
recall that $g(\x)=\ds\frac{\partial u(\x)}{\p\nu_{\x}}$ and use in
(\ref{eq:comp}) the approximation
 \beq
g(\x)\sim
g_0(\x)+g_1(\x)\sim\ds\frac{-2C}{a\pi\sqrt{1-\ds\frac{|\x|^2}{a^2}}}\left[1+\frac{L(\mb{0})+
N(\mb{0})}{2\pi}\,a\log a\right].
 \eeq
We obtain the narrow escape time $E\tau=C$ (in dimensionless
variables) as
 \beq
 E\tau=\frac{|\Omega|}{4aD\left[1+\ds\frac{L(\mb{0})+
N(\mb{0})}{2\pi}\,a\log a+o(a\log a)\right]}.
 \eeq
The principal eigenvalue $\lambda_1(a)\sim\ds\frac1{E\tau}$ of the
Laplace equation in $\Omega$ with the  mixed Dirichlet-Neumann
boundary conditions (\ref{eq:boundary-condition-v}), (\ref{dvdn})
has the asymptotic expansion for $\eps\to0$
 \beq
 \lambda_1(a)=\frac{4aD}{|\Omega|}\left[1+\ds\frac{L(\mb{0})+
N(\mb{0})}{2\pi}\,a\log a+o(a\log a)\right].\label{lambda1}
 \eeq
The result (\ref{lambda1}) provides the missing second term and
estimate of the remainder, which was not given in
\cite{Ward1,Ward2,Ward3,Ward4,SSH1}.

If $\Omega$ is a ball of radius $R$, then $L(\mb{0})+
N(\mb{0})=\ds\frac2R$ and the narrow escape time $E\tau=C$ is given
(in dimensional variables) by
 \beq
 E\tau=\frac{|\Omega|}{4aD\left[1-\ds\frac{a}{\pi R}\,\log \ds\frac{R}{a}+
 o\left(\ds\frac{a}{R}\log\ds\frac{R}{a}\right)\right]}=\ds{\frac{|\Omega|}{4aD}} \left[1+\frac{a}{\pi R}
\log \frac{R}{a} + o\left(\frac{a}{R}\log\frac{R}{a}\right)
\right].\label{Etau}
 \eeq
The result (\ref{Etau}) corrects that given in \cite{SSH1}.
Specifically, equation (3.52) in \cite{SSH1} is missing the factor
$1/\pi$ of equation (\ref{Etau}), which should have been carried
from eq.(3.51) in \cite{SSH1}. The case of an elliptic window is
handled in a straightforward manner, as in \cite{SSH1}.

\subsection{Leakage in a conductor of Brownian particles}

A conductor of Brownian particles is a bounded domain $\Omega$, with
a source of particles on the boundary or in the interior and a
target, which is an absorbing part $\p\Omega_a$ of $\p\Omega$. The
remaining boundary $\p\Omega_r$ is reflecting. Some of the Brownian
particles may leak out of $\Omega$  if $\p\Omega_r$ contains a small
absorbing hole {\bf $S(\varepsilon)$}. The calculation of the
leakage flux is not the same as that in the narrow escape problem,
because the total flux on the boundary remains bounded as the small
hole shrinks. Our purpose is to find the portion that leaks through
the small hole out of the total flux.

The (dimensionless) stationary density $u(\x)$ of the Brownian
particles satisfies the mixed boundary value problem
 \beq \label{dynb}
 D\Delta u(\x)&=&0\quad\mbox{for}\quad
\x\in\Omega\nonumber \\
& &\nonumber\\
\left.\frac{\p u(\x)}{\p\nu}\right|_{\p\Omega_r}
&=&0\nonumber\\
& & \\
- D\left.\frac{\p u(\x)}{\p\nu}\right|_{\p \Omega_s}
& =&\phi(\x)\nonumber\\
& &\nonumber \\
\left.u(\x)\right|_{\p\Omega_a}&=& \left.u(\x)\right|_{ S(\eps)}=
0,\ \nonumber
  \eeq
where $\phi(\x)$ is the flux density of the source on the boundary.
Next, we derive an asymptotic expression for the flux through
$S(\eps)$,
 \beq
 J_{\eps}= D\int_{ S(\eps)}\frac{\p
u(\x)}{\p\nu}\,dS_{\x},
 \eeq
in terms of the solution $u_0(\x)$ of the reduced problem (without
$S(\eps)$), thus avoiding the need to construct boundary layers.
First, we find the flux of each eigenfunction and then, using
eigenfunction expansion, we calculate $ J_{\eps}$. Every
eigenfunction $u_{\eps}(\x)$ of the homogeneous problem (\ref{dynb})
satisfies
 \beq
- D\Delta u_{\eps}(\x) &=&\lambda(\eps) u_{\eps}(\x)\quad
\mbox{for}\quad\x\in\Omega\label{efeq} \\
&&\nonumber\\
\ds\frac{\p u_{\eps}(\x)}{\p \nu}&=& 0\quad\hbox{for}\quad
\x \in \p \Omega_s\cup\p\Omega_r\label{dynb1}\\
&&\nonumber\\
u_{\eps}(\x)&=& 0\quad \hbox{for}\quad \x \in  S(\eps)\cup
\p\Omega_a\label{u1bc}.
 \eeq
The matched asymptotics method of \cite{Ward1}-\cite{Ward4} gives
the expansion of the eigenvalues
 \beq
 \lambda(\eps) = \lambda(0)+\lambda_1 \eps+
o(\eps),
 \eeq
where $\lambda(0)$ is the eigenvalue of the reduced problem (for
$\Omega$ without any small holes).

We define the reduced Green function (without the small hole) as the
solution of the mixed boundary value problem with $ D=1$,
 \beq
-\Delta G(\x,\y) &=& \delta(\x-\y)
\quad\mbox{for}\quad\x,\y\in\Omega \label{Green}\\
&&\nonumber\\
\ds\frac{\p G}{\p \nu}(\x,\y)&=& 0\quad\hbox{for}\quad
\x \in \p\Omega_s\cup \Omega_{r},\ \y\in\Omega\label{dGnu}\\
&&\nonumber\\
G(\x,\y)&=& 0, \quad\hbox{for}\quad \x \in \p\Omega_a,\
\y\in\Omega\label{Gabsorb}.
 \eeq

Multiplying (\ref{Green}) by $u_\eps(\y)$ and integrating over
$\Omega$, we get
 \beq\label{equ1}
u_{\eps}(\x) = \frac{\lambda(\eps)}{D} \int_{\Omega} G(\x,\y) u_\eps(\y)\,
d\y+ \int_{ S(\eps)} G(\x,\y) \frac{\p u_{\eps}(\y)}{\p\nu}\,
dS_{\y}.
 \eeq
In view of the boundary condition (\ref{u1bc}), we get from
(\ref{equ1}) for all $\x \in S(\eps)$
 \beq
  \label{eqf1} \frac{\lambda(\eps)}{D}
\int_{\Omega} G(\x,\y) u_\eps(\y)\, d{\y}=-\int_{ S(\eps)}
G(\x,\y)\frac{\p u_{\eps}(\y)}{\p \nu}\, dS_{\y}.
 \eeq
The integral on the left hand side of (\ref{eqf1}) can be expanded
about the center of $ S(\eps)$ in the form
 \beq
\int_{\Omega} \lambda(\eps) G(\x,\y) u_{\eps}(\y)\, d{\y}=
 G_0(\eps)+O(|\x|)\quad\mbox{for}\quad\x\in
  S(\eps),\label{powers}
 \eeq
where the origin is assumed to be in the center of $ S(\eps)$ and
the $(x_1,x_2)$ plane is that of $ S(\eps)$.

As in Section \ref{ss:small-circular}, Green's function for the
mixed boundary value problem has the form
 \begin{equation}
\label{eq:Neumann-v} G(\x ,\y) = \frac{1}{2\pi |\x -\y|} +
H(\x,\y)\log|\x-\y|+v_S(\x ,\y),
\end{equation}
for $\x\in\p\Omega,\ \y\in\Omega\cup\p\Omega$, where $H(\x,\y)$
depends locally on the curvatures of the boundary and $v_S(\x,\y)$
is a continuous function of $\x,\y\in\Omega$ and on $\p\Omega$. We
assume that $H(\x,\y)$ is bounded. Using (\ref{eq:Neumann-v}) and
the expansion (\ref{powers}) in (\ref{eqf1}), we obtain the
Helmholtz equation
 \beq
\frac{ G_0(\eps)}{D}+O(|\x|)=-\int_{S(\eps)}\left[\frac{1}{2\pi |\x
-\y|} + H(\x,\y)\log|\x-\y|+v_S(\x ,\y)\right]\ds\frac{\p
u_{\eps}(\y)}{\p \nu}\,dS_{\y}.
 \eeq
The leading order singularity of $G(\x,\y)$ and (\ref{intg0})
suggest the expansion
 \beq
\ds\frac{\p u_{\eps}(\y)}{\p \nu}=
\ds\frac{C_0(\eps)}{\sqrt{1-\ds\frac{|\y|^2}{\eps^2}}}+O(|\y|)
\quad\mbox{for}\quad\y\in S(\eps),\label{gs}
 \eeq
where $C_0(\eps)$ is yet an undetermined coefficient, that is,
 \beq
\frac{G_0(\eps)}{D}+O(|\x|)&=&-\int_{ S(\eps)}\left[\frac{1}
{2\pi|\x-\y|}+ H(\x,\y)\log|\x-\y|+v_S(\x,\y)\right]\times\nonumber\\
&&\nonumber\\
&&\left[
\ds\frac{C_0(\eps)}{\sqrt{1-\ds\frac{|\y|^2}{\eps^2}}}+O(|\y|)\right]\,dS_{\y}\label{Helm}
 \eeq
which reduces at $\x=\mb{0}$ to
 \beqq
\frac{G_0(\eps)}{D}&=&\frac{-C_0(\eps)\pi\eps }{2}+\int_{ S(\eps)}
O(|\y|)\left[\frac{1}{2\pi|\y|}+H(0,\y)
\log|\y|\right]\,dS_{\y}-\nonumber\\
&&\nonumber\\
&&\nonumber\\
&&\int_{
S(\eps)}\frac{C_0(\eps)\left[H(\mb{0},\y)\log|\y|+v_S(\mb{0},\y)
\right]\,dS_{\y}} {\sqrt{1-\ds\frac{|\y|^2}{\eps^2}}}+\int_{
S(\eps)}O(|\y|)\,dS_{\y}.
 \eeqq
It follows that
 \beqq
 \frac{G_0(\eps)}{D}=-\left(\frac{\pi\eps}{2}+O(\eps^2\log\eps)\right)C_0(\eps)
 +O(\eps^2\log\eps),
 \eeqq
so that
 \beq
 C_0(\eps)=-\frac{G_0(\eps)+O(\eps^2\log\eps)}{D\left[\ds\frac{\pi\eps}{2}
 +O(\eps^2\log\eps)\right]}.\label{C0e}
 \eeq
Now, (\ref{gs}) gives the flux through $S(\eps)$ as
 \beq
  -D\int_{ S(\eps)} \ds\frac{\p u_{\eps}(\y)}{\p
\nu}\,dS_{\y}&=& \frac{G_0(\eps)+O(\eps^2\log\eps)}
{\ds\frac{\pi\eps}{2}+O(\eps^2\log\eps)} \int_{
S(\eps)}\frac{dS_{\y}}{\sqrt{1-\ds\frac{|\y|^2}{\eps^2}}}
+\nonumber\\
&&\nonumber\\
&& \int_{ S(\eps)} O(|\y|)\,dS_{\y}\nonumber\\
&&\nonumber\\
&=&4\eps \frac{G_0(\eps)+O(\eps^2\log\eps)}{1+O(\eps\log\eps)}+
O(\eps^2\log\eps)\label{Du1}.
 \eeq
To determine $G_0(\eps)$, we integrate (\ref{efeq}), to get the
total flux condition
 \beq
\label{eqf2} \lambda(\eps)\int_{\Omega} u_{\eps}(\x)\,d\x = D\int_{
S(\eps)} \ds\frac{\p u_{\eps}(\y)}{\p \nu}\,dS_{\y}+
D\int_{\p\Omega_a} \ds\frac{\p u_{\eps}(\y)}{\p \nu}\,dS_{\y}.
 \eeq
We also recall that (\ref{Gabsorb}) implies that
 \beqq
\int_{\p \Omega_a} G(\x,\y)\frac{\p u_{\eps}(\y)}{\p \nu}\, dS_{\y}
=0\quad\mbox{for}\quad\x\in\p\Omega_a,
 \eeqq
hence, using equations (\ref{eqf1}) and (\ref{eqf2}), we get the two
equations
 \beq
\lambda(\eps)\int_{\Omega} u_{\eps}(\x)\,d\x& =&4\eps
\frac{G_0(\eps)+O(\eps^2\log\eps)}{1+O(\eps\log\eps)}+O(\eps^2\log\eps)
\nonumber\\
&&\nonumber\\
&&\nonumber\\
&&+ D\int_{\p\Omega_a}\ds\frac{\p u_{\eps}(\y)}{\p
\nu}\,dS_{\y} \nonumber\\
&&\nonumber\\
\lambda(\eps)  \int_{\Omega} G(\mb{0},\y) u_{\eps}(\y)\, d{\y}&=&
G_0(\eps).\label{G0eps}
 \eeq
This gives
 \beqq
&&\lambda(\eps)\int_{\Omega} u_{\eps}(\x)\,d\x =\nonumber\\
&&\nonumber\\
&&  \frac{4\eps\lambda(\eps) \ds\int_{\Omega} G(\mb{0},\y)
u_{\eps}(\y)\,
d{\y}+O(\eps^2\log\eps)}{1+O(\eps\log\eps)}+O(\eps^2\log\eps)+
D\int_{ \p\Omega_a}\ds\frac{\p u_{\eps}(\y)}{\p \nu}\,dS_{\y}.
 \eeqq
Solving for $\lambda(\eps)$, we find that
 \beq
\lambda(\eps)&=&\frac{ D\ds\int_{ \p\Omega_a}\ds\frac{\p
u_{\eps}(\y)}{\p \nu}\,dS_{\y}+O(\eps^2\log\eps)}{\ds\int_{\Omega}
u_{\eps}(\x)\,d\x-\ds\frac{4\eps }{1+O(\eps\log\eps)}
\ds\int_{\Omega} G(\mb{0},\y) u_{\eps}(\y)\,
d{\y}+O(\eps^2\log\eps)}\nonumber\\
&&\nonumber\\&&\nonumber\\
&=&\frac{ D\ds\int_{ \p\Omega_a}\ds\frac{\p u_{\eps}(\y)}{\p
\nu}\,dS_{\y}}{\ds\int_{\Omega}
u_{\eps}(\x)\,d\x}\left(1+\ds\frac{4\eps \ds\int_{\Omega}
G(\mb{0},\y) u_{\eps}(\y)\, d{\y}}{\ds\int_{\Omega}
u_{\eps}(\x)\,d\x}\right)+O(\eps^2\log\eps).\label{lambdae}
 \eeq
Note that
 \beq
\frac{ D\ds\int_{ \p\Omega_a}\ds\frac{\p u_{\eps}(\y)}{\p
\nu}\,dS_{\y}}{\ds\int_{\Omega}
u_{\eps}(\x)\,d\x}=\lambda(0)+O(\eps),\label{lambda0}
 \eeq
due to the contribution of the boundary layer near $ S(\eps)$.

Obviously, $u_{\eps}\to u_0$ as $\eps\to0$, where $u_0$ is the
corresponding eigenfunction of the reduced problem (in the absence
of the small hole, see also \cite{Ward1}), so
 \beqq
\lim_{\eps\to0}\int_{\Omega} G(\x,\y) u_{\eps}(\y)\, d{\y} =
\int_{\Omega} G(\x,\y) u_0(\y)\,d{\y}, \quad \lim_{\eps\to0}
\int_{\p\Omega_a} \ds\frac{\p u_{\eps}(\y)}{\p \nu}dS_{\y}
=\int_{\p\Omega_a} \ds\frac{\p u_0(\y)}{\p \nu}\,dS_{\y}.
 \eeqq
Therefore, using (\ref{G0eps})-(\ref{lambda0}) in (\ref{Du1}), we
find that the flux of $u_{\eps}(\x)$ through the small hole is
 \beq
 J(\eps)&=& -D\int_{ S(\eps)} \ds\frac{\p
u_{\eps}(\y)}{\p \nu}\,dS_{\y}=4\eps\lambda(0) \int_{\Omega}
G(\mb{0},\y) u_0(\y)\,
d{\y}+O(\eps^2\log\eps)\nonumber\\
&&\nonumber\\
&=&4\eps Du_0(\mb{0})+O(\eps^2\log\eps).\label{Jeps}
 \eeq
Finally, expanding the solution $u(\x)$ of (\ref{dynb}) in
eigenfunctions, we obtain from (\ref{Jeps})
 \beq
 J_{\eps}=4\eps
Du_0(\mb{0})\left(1+O(\eps\log\eps)\right),\label{Jeps4}
 \eeq
where $u_0(\mb{x})$ is the solution of the reduced problem
(\ref{dynb}). In dimensional variables, we obtain
 \beq
J_{\eps}=4a
Dp_0(\mb{0})+O\left(\frac{a^2}{|\Omega|^{2/3}}\log\frac{a}{|\Omega|^{1/3}}\right),\label{Jhole}
 \eeq
where $p_0(\mb{0})$ is the value of the reduced stationary density
(without the perforation) at the hole.

\section{Summary and discussion}

The main results of this paper are (i) the explicit calculation of
the second term in the expansion of the NET, which can be quite
significant, and which also provides a bound for the remainder in
the expansion; (ii) an explicit expression for the leakage flux
through a small opening in the impermeable envelope of a conductor
of ions. The leakage is often a key control mechanism of
physiological function, such as in the synaptic cleft of a neuron,
as mentioned in the Introduction.  The leakage formula (\ref{Jhole})
can give explicit expressions for the flux when the reduced problem
is explicitly solvable, e.g., in simple geometries. If there are
several leaks, at $\x_i$, then (\ref{Jhole}) gives
 \beq
J_{\eps}=4a
D\sum_ip_0(\x_i)+O\left(\frac{a^2}{|\Omega|^{2/3}}\log\frac{a}{|\Omega|^{1/3}}\right),\label{Jholei}
 \eeq
which demonstrates the role of clustering or un-clustering of the
leaks in regulating flux \cite{PLA-holes}, \cite{JPA-holes}.
Specific applications of the results of this paper to molecular
biology and biophysics will be published in a separate paper.

\begin{acknowledgments}
AS thanks the Yale University-Weizmann Institute Joint Research Fund; ZS was partially supported by
a research grant from TAU; DH was partially supported by an ERC-starting grant and an HFSP
research grant.
\end{acknowledgments}

\end{document}